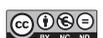

Original Research Article

# Co-creation and evaluation of an app to support reminiscence therapy interventions for older people with dementia

Iván De-Rosende-Celeiro[1] 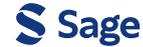, Virginia Francisco-Gilmartín[2],
Susana Bautista-Blasco[3] and Adriana Ávila-Álvarez[4]

## Abstract

**Objective:** The objectives encompassed (1) the creation of *Recuerdame*, a digital app specifically designed for occupational therapists, aiming to support these professionals in the processes of planning, organizing, developing, and documenting reminiscence therapies for older people with dementia, and (2) the evaluation of the designed prototype through a participatory and user-centered design approach, exploring the perceptions of end-users.

**Methods:** This exploratory research used a mixed-methods design. The app was developed in two phases. In the first phase, the research team identified the requirements and designed a prototype. In the second phase, experienced occupational therapists evaluated the prototype.

**Results:** The research team determined the app's required functionalities, grouped into eight major themes: register related persons and caregivers; record the patient's life story memories; prepare a reminiscence therapy session; conduct a session; end a session; assess the patient; automatically generate a life story; other requirements. The first phase ended with the development of a prototype. In the second phase, eight occupational therapists performed a series of tasks using all the application's functionalities. Most of these tasks were very easy (Single Ease Question). The level of usability was considered excellent (System Usability Scale). Participants believed that the app would save practitioners time, enrich therapy sessions and improve their effectiveness. The qualitative results were summarized in two broad themes: (a) acceptability of the app; and (b) areas for improvement.

**Conclusions:** Participating occupational therapists generally agreed that the co-designed app appears to be a versatile tool that empowers these professionals to manage reminiscence interventions.

Submission date: 31 July 2023; Acceptance date: 29 May 2024

## Introduction

Dementia is a neurodegenerative disorder that is becoming a major and increasingly prevalent public health and social problem around the world.[1] In 2019, it was estimated that some 55 million people worldwide were living with dementia.[2] This neurocognitive disorder gradually reduces functions such as memory, communication or problem-solving skills, severely limiting the ability to perform everyday

[1]Facultad de Ciencias de la Salud, Universidade da Coruña, A Coruña, Spain
[2]Departamento de Ingeniería del Software e Inteligencia Artificial, Facultad de Informática, Universidad Complutense de Madrid, Madrid, Spain
[3]Escuela Politécnica Superior, Universidad Francisco de Vitoria, Madrid, Spain
[4]Facultad de Ciencias de la Salud, Universidade da Coruña, A Coruña, Spain

**Corresponding author:**
Iván De-Rosende-Celeiro, Facultad de Ciencias de la Salud, Universidade da Coruña, Campus Universitario de Oza, s/n, 15006, A Coruña, Spain.
Email: ivan.de.rosende.celeiro@udc.es

tasks. Researchers are turning to non-pharmacological approaches to enhance the well-being and quality of life of the affected population.[3] Psychosocial interventions by professionals such as occupational therapists, increasingly use reminiscence and life story work therapies in treating dementia.[4] These consist of narrating and discussing past experiences in one's own life, with another person or in group sessions.[5] Typically, people with dementia are better able to recall episodes from their earlier life, especially from childhood and early adulthood, than more recent events.[6] Reminiscence therapy takes advantage of this and stimulates cognitive activity through a strengths-based approach, thus promoting engagement and communication in a climate of greater self-confidence.[4,7] Therapists use a wide variety of memory triggers and tangible prompts, such as photographs, objects from the past, historical materials, music, or recordings.[8]

Several meta-analyses and reviews of scientific literature show that the use of psychosocial interventions based on reminiscence and life story therapies can be effective. A 2022 meta-analysis of 29 randomized controlled trials or quasi-experimental studies into the effectiveness of these therapies found improvements in cognitive function, quality of life, depressive and neuropsychiatric symptoms.[9] The effect sizes for each outcome ranged from small to moderate.[9] Another meta-analysis suggested that reminiscence work may be an alternative to pharmacological treatment in treating depression among people with dementia.[10] The authors evaluated 24 randomized controlled clinical trials, identifying a medium effect size on depressive symptomologies and improvements in aspects such as quality of life or behavioral symptoms.[10] Similarly, the Cochrane meta-analysis concluded that reminiscence and life story therapies can improve quality of life, cognition, communication, and mood.[4] Finally, a systematic review of studies on the efficacy of life story books found that this therapy had positive effects on the autobiographical memory and depression among those with dementia, improving the quality of relationships with caregivers and mitigating the burden on informal caregivers.[11]

More recently, there has been growing interest in the use of digital tools as a mean to facilitate reminiscence strategies, using information and communication technologies (ICT) such as multimedia mobile applications, computer-assisted programs or digital memory books.[12,13] However, the availability of technology-aided applications continues to be very limited. A 2023 meta-analysis of ICT-based interventions for people with dementia found only two randomized controlled experimental studies dealing with reminiscence and life story therapies.[14] In two systematic reviews into the use of digital technologies, published in 2020 and 2021, the small number of studies led the authors to recommend that the development of multimedia tools to support these therapies should be a top research priority.[8,15] Significantly, none of the studies analyzed in the Cochrane meta-analysis used ICT innovations during therapeutic interventions.[4]

Against this backdrop, our study addresses the gap in research on ICT tools for reminiscence and life story therapies in treating dementia. We present *Recuerdame*, a digital app tailored for occupational therapists, with the aim of simplifying and optimizing their workflow. Our hypothesis suggests that *Recuerdame* can enhance therapeutic processes, supporting professionals in planning, organizing, developing, and documenting reminiscence sessions. The research objectives encompassed the creation of *Recuerdame* and the evaluation of the prototype through a participatory and user-centered design approach, exploring the perceptions of end-users.

## Methods

### Study design

This research project used a mixed-methods design. Qualitative and quantitative information was collected between January and June 2023 in order to draw general conclusions about the perspectives of end-users. The qualitative research applied the conventional content analysis method with an exploratory approach.

### Procedures and data collection

The app was developed in two phases. In the first phase, the research team conducted an analysis to identify the requirements for the app and designed a prototype. In the second phase, an exploratory study of the prototype was carried out from the perspective of end-users. According to the principles of a participatory and user-centered approach,[16,17] characterized by the active involvement of future users in the app development process (co-creation), considering the user perspective from the design stage, a sample of experienced occupational therapists evaluated the prototype.

*Initial prototype design phase.* In the first phase, the members of the research team and the developers worked together to design an initial prototype of the application. *Recuerdame* is part of a larger funded research project (CANTOR) that included a multidisciplinary research team consisting of four experts with PhDs in computer science (Complutense University of Madrid, Spain) and occupational therapy (University of A Coruna, Spain). The two experts in computer science, with an average of 10 years of research experience in the field, conducted all the interviews included in the study. The interviewers had previous experience in conducting interviews. The two expert occupational therapists, with an average of 22 years of research experience, described their general technology proficiency as "medium". Four developers, all graduates in computer science, also participated in the project.

The occupational therapists of the research team determined the overall concept, User Interface (UI) and User Experience (UX) of the app. A semi-structured discovery interview was conducted with these two therapists of the research team to identify the required functionalities of the app. In the interview the therapists told of their own experience using reminiscence therapy with people living with dementia.[18] An interview schedule for predetermined topics was created, taking available literature as a reference,[12,13] but allowed for free-flowing responses and ideas by using open-ended questions (Table 1).

Once the required functionalities were decided, the two experts in computer science of the research team and the developers created a prototype. This was then presented to the two occupational therapists of the research team, who were asked to perform a series of tasks (described in Table 2) using all the app functionalities. After completing each task, these occupational therapists responded to the Single Ease Question (SEQ)[19–21] rating the difficulty of the task using a 7-point Likert-type scale (1: very difficult to 7: very easy). The question was: "How difficult was it to do this task altogether?". The mean SEQ response was from 5.3 to 5.6 points. This benchmark helped to determine which tasks are particularly easy or difficult for the evaluators, with a view to making improvements. Once the final design was agreed upon by the research team, the developers proceeded to create a functional prototype.

*Evaluation phase.* After a functional prototype of the app was developed, it was evaluated using theoretical sampling with a group of experienced occupational therapists, none of them belonging to the research team.

*Participants and recruitment.* Participants in the second phase were selected based on their ability to provide input relevant to the research objectives. The study population consisted of occupational therapists with extensive experience in the field of therapeutic care for older people with dementia. The inclusion criteria for the theoretical sampling

**Table 1.** Semi-structured interviews: predetermined topics and open-ended questions.

| Initial prototype design phase interview |
|---|
| Reminiscence therapy<br>- What is reminiscence therapy?<br>- What is this type of therapy used for? |
| Life story<br>- What is a life story?<br>- What do these life stories include (audio, text, images…)?<br>- How should these life stories be shown to be useful in their therapies? |
| Workflow of the app to be designed<br>- What would you need from a computer app to help you with reminiscence-based therapies?<br>- How would you use an app that showed you life stories?<br>- How would you like the memories to be ordered in the life story? By date? Alphabetically? By labels (family, friends, hobbies, travel…)?<br>- Would you need to perform searches in the app? What data? Which criteria?<br>- Would you need to do some kind of monitoring of the patient's evolution? What data should be retained from one session to another?<br>- Do you have any further comments on the digital app to be designed? |
| Evaluation phase interview |
| - What did you think of the app?<br>- Is there any feature of this app that you particularly like?<br>- Is there any feature of this app that you particularly dislike?<br>- Is there anything you would change about this app? Please specify the changes you would make<br>- Do you have any further comments about this app? |

**Table 2.** Initial prototype design phase: Single Ease Question questionnaire ($n=2$)

| Sequence | Tasks | Mean |
|---|---|---|
| 1 | Access to a patient | 7 |
| 2 | Create a reminiscence session | 7 |
| 3 | End a reminiscence session and create the follow-up report | 4.5 |
| 4 | Perform a clinical assessment of a patient | 7 |
| 5 | Modify a session in a patient's calendar | 7 |
| 6 | Automatically generate a life story | 7 |
| 7 | Identify related persons from a specific memory | 6.5 |
| 8 | Add an existing media file to a memory | 6 |
| 9 | Add a related person | 7 |
| 10 | Write an email to a patient | 4.5 |
| 11 | Sign off | 7 |

Items are scored on a 7-point Likert-type scale from 1 (very difficult) to 7 (very easy).

were: (1) currently a practicing professional in occupational therapy in the field of dementia; and (2) a minimum of five years of experience as an occupational therapist in this field. The sample was drawn from only within the Spanish province of A Coruña, for reasons of accessibility and sample control. A professor of the Occupational Therapy at the University of A Coruña, with extensive knowledge of the network of care for older people in the province, proposed an initial list of key informants. These proposed informants were initially approached via email by the research team, who explained the scope and objectives of the research and requested their participation. The final sample size was reached by saturation; that is, when conducting a new interview did not provide new data.[22]

In this second phase, all the invited therapists agreed to participate. The saturation point of the discourse was reached in the eight interviews, given the homogeneity of the recruited profile. Therefore, eight experienced therapists were recruited for the assessment of the functional prototype. The mean age of the sample was 35.4 years (SD 2.03). 87.5% of the participants were women. All participants applied therapeutic interventions based on reminiscence and life story therapy within their professional practice in care institutions for older people with dementia. One participant was also a psychologist. Four therapists worked in a day care center, three in a nursing home, and one worked with an association. Their experience in the use of reminiscence therapy averaged 8.25 years (median, Q1—Q3 = 5.5–10, range 5–20) while their level of technical proficiency was self-assessed as "medium" by six therapists, "high" by one, and "low" by another participant.

*Procedures.* The evaluation of the functional prototype was conducted by the computer science experts within the research team. Individual evaluators accessed *Recuerdame* as a web application, utilizing their personal computers and accessing it through their installed browsers. On average, the evaluation process lasted approximately 63 min, with each session being conducted independently and featuring a single evaluator. These evaluation sessions were facilitated through Google Meet, with the computer science experts conducting the sessions. Only the researcher and the participant were present during the sessions. Evaluators shared their screens to enable expert observation of their interactions with the *Recuerdame* app throughout the process. Brief field notes were made during the evaluation session.

The videoconference sessions began with a demo of the app by the research team to provide participants with a general overview of how it worked. Subsequently, occupational therapists engaged in a series of tasks using *Recuerdame* (described in Table 3), covering all aspects of the application's functionalities. Following the completion of each task, evaluators responded to the SEQ. Once the entire tasks set was concluded, testers evaluated the prototype's usability using the Spanish version of the System Usability Scale (SUS), a well-established tool for assessing system usability.[23] This version proved to be valid and reliable for assessing the usability of electronic tools in healthcare settings.[23] The SUS consists of ten statements (Table 4) rated on a 5-point Likert-type scale, from 1 (strongly disagree) to 5 (strongly agree). Of the ten statements, five are positive (1, 3, 5, 7, and 9) and five are negative (2, 4, 6, 8, and 10). The total score ranges from 0 to 100. Scores above 68 indicate above-average usability, while scores exceeding 85 indicate excellent usability.[23]

Therapists were then given an ad hoc quantitative questionnaire to measure their perceptions of the potential benefits of using the app in reminiscence/life story therapy. The questionnaire, consisting of 8 statements (Table 5), was meticulously crafted with questions grouped into the domains of 'time savings', 'enrichment of therapy', and 'improvement of effectiveness'. The Likert-type scale used for responses ranged from 1 (strongly disagree) to 5 (strongly agree). Prior to its implementation, the questionnaire underwent validation through pilot testing, ensuring clarity, relevance, and effectiveness. Adjustments were made based on feedback, and the final questionnaire was administered during evaluation sessions through Google Forms for convenience and accessibility, following comprehensive testing of the online platform.

The evaluation sessions concluded with a semi-structured interview, allowing participants to expand on their views. To gather more in-depth qualitative information on the perceived value of the app, participants responded to a set of open-ended questions (see Table 1).

### Ethical statement

This research was approved by the Research Ethics Committee of the University of A Coruña. All participants were fully informed, orally and in writing, about the research objectives and procedures, including video and audio recording, and signed an informed consent form before participating in the study. The research adhered to the Declaration of Helsinki.[24] Data was securely stored to safeguard confidentiality, in accordance with the European Union General Data Protection Regulation.

### Analysis

The quantitative data was analyzed using descriptive statistics. The categorical variables were reported as frequencies and percentages. The Shapiro–Wilk test was used to determine the normal distribution. The variables that followed a normal distribution were described using the mean and the standard deviation (SD); those that did not follow the normal distribution and the ordinal variables, were described using the median and the first and third quartiles

Table 3. Evaluation phase: single ease question questionnaire (n = 8).

| Sequence | Tasks | Median | Q1–Q3 |
| --- | --- | --- | --- |
| 1 | *Reminiscence sessions and memories*<br>Select a patient, review the list of sessions and identify which sessions were completed. Select a session and review its data. Select a memory and review its data. Detect needs for improvement | 7 | 6–7 |
| 2 | *Reminiscence session reports*<br>Search the available session reports. Select a session report, review its data, detect needs for improvement, and generate a PDF report | 6.5 | 6–7 |
| 3 | *Clinical assessment reports*<br>Search available clinical assessment reports. Select a report, review its data, detect needs for improvement, and generate a PDF report | 7 | 6.25–7 |
| 4 | *Automatic generation of the life story*<br>Generate the life story of a patient with memories of a specific stage of the life and of a specific time interval. Generate a life story book, interact with the photo carousel, and identify needs for improvement. Generate a PDF document, review its data, and detect improvement needs | 7 | 6–7 |
| 5 | *Edit memories*<br>Consult the list of memories, select a specific memory, review its data, and detect improvement needs. Modify a given memory and add materials such as images | 6 | 6–6 |
| 6 | *People related to the patient*<br>Search for people related to a patient, select a particular person, review available data on that person, determine improvement needs, and sort the list of related persons by relationship type | 7 | 5.25–7 |
| 7 | *Patient calendar with schedule of sessions*<br>Review the calendar with the list of sessions scheduled for the patient. Select a given session, review its data, and modify it. Create a new session | 6 | 6–7 |
| 8 | *Create and edit a reminiscence session*<br>Create a new session, add a memory, add an image, and save the session. Modify the created session, end it, and generate the session report | 6 | 5–6 |
| 9 | *Create and edit a clinical assessment*<br>Create a new clinical assessment of a patient, save it, and modify it | 7 | 6.25–7 |
| 10 | *Patient registration*<br>Add a new patient in the app | 7 | 7–7 |
| 11 | *Caregiver registration*<br>Register a new caregiver in the app and link him/her to a patient | 6.5 | 6–7 |

Items are scored on a 7-point Likert-type scale from 1 (very difficult) to 7 (very easy).
Q1, first quartile; Q3, third quartile.

(Q1–Q3). Statistical analyses were conducted using IBM SPSS 25.0.

All interview sessions were audio-recorded and transcribed verbatim by trained members of the research team who were present during the evaluation sessions. To ensure accuracy, recordings were conscientiously played back during the transcription process, and transcripts were cross-checked against the audio recordings to identify and correct any discrepancies. Transcripts were not returned to participants for comment. Responses to the open questions were analyzed using thematic analysis, identifying, analyzing, and reporting patterns within the data.[25,26] The content was analyzed in a classical way.[27] The themes of analysis were extracted in a mixed way, from the interview guides and those emerging from the data. Each sentence or paragraph was assigned codes summarizing its meaning which were then grouped according to their similarity. After identifying these patterns, key descriptive themes and sub-themes were formed. In the co-creation phase, this analysis was carried out by the two computer

Table 4. Evaluation phase: system usability scale (n = 8).

| No. | Items | Median | Q1–Q3 |
| --- | --- | --- | --- |
| 1 | I think that I would like to use this system frequently | 5 | 4–5 |
| 2 | I found the system unnecessarily complex | 1 | 1–1.75 |
| 3 | I thought the system was easy to use | 5 | 4–5 |
| 4 | I think that I would need the support of a technical person to be able to use this system | 1 | 1–2 |
| 5 | I found the various functions in this system were well integrated | 4.5 | 4–5 |
| 6 | I thought there was too much inconsistency in this system | 1.5 | 1–2 |
| 7 | I would imagine that most people would learn to use this system very quickly | 5 | 4–5 |
| 8 | I found the system very cumbersome to use | 1 | 1–1.75 |
| 9 | I felt very confident using the system | 5 | 4.25–5 |
| 10 | I needed to learn a lot of things before I could get going with this system | 1 | 1–2 |

Items are scored on a 5-point Likert-type scale from 1 (strongly disagree) to 5 (strongly agree).
Q1, first quartile; Q3, third quartile.

Table 5. Evaluation phase: benefits of using the app in reminiscence and life story interventions (n = 8).

| Items | Median | Q1–Q3 |
| --- | --- | --- |
| I believe that the use of the app would reduce the time needed to record information about the person's memories and biography | 4 | 4–5 |
| I believe that the use of the app would allow me to save time in the preparation of therapy sessions | 4.5 | 4–5 |
| I believe that the use of the app would reduce the time I spend locating memories and biographical materials of the subject relevant to therapy | 5 | 4.25–5 |
| I believe that the use of the app would allow me to save time in the task of recording information about the development of the therapy sessions | 4 | 4–4.75 |
| I believe that the use of the app would reduce the time I spend tracking the person's progress | 4 | 4–5 |
| I believe that the use of the app would allow me to streamline the reminiscence and life story based therapies | 4 | 4–4.75 |
| I believe that using the app to obtain multimedia narratives of life stories would enrich therapy | 5 | 4.25–5 |
| I believe that the use of the app during my therapy sessions would make them more effective | 5 | 4–5 |

Items are scored on a 5-point Likert-type scale from 1 (strongly disagree) to 5 (strongly agree).
Q1, first quartile; Q3, third quartile.

science experts; in the evaluation phase, the final themes were decided by consensus among the four members of the research team. Any discrepancies were discussed. Through discussions, a consensus on the underlying themes was agreed upon and those results are presented. Textual quotations were selected to illustrate the sub-themes and all quotes were translated into English. The 32-item Consolidated Criteria for Reporting Qualitative Studies (COREQ) checklist[28] was used to ensure standardization of reporting and rigor (Supplementary Appendix 1).

## Results

### Initial prototype design phase

*App requirements.* The semi-structured interviews with the therapists of the research team to determine the app's required functionalities revealed that *Recuerdame* should allow the practitioner to perform a set of specific tasks. These are described below, grouped into eight major themes.

*Register related persons and caregivers.* The app should allow the therapist to register people involved in the patient's life history and their caregivers. The following information may be added: name, type of relationship, and contact information. The digital tool should allow the practitioner to send emails to these registered persons.

*Record the patient's life story memories.* It should allow the therapist to record the patient's life story memories at any time. This record should include information such as a description of the memory, location, date, type (childhood, adolescence, young adult, adult, and older adult), category (e.g., family, friends, work, hobbies, and pets), people related to the memory (e.g., caregivers or relatives), and preservation status (memory preserved, at risk of loss, or lost). The practitioner should be able to record the emotion the memory evokes in the patient and label it as positive-neutral-negative, as well as the degree to which the memory produces a positive mood on a quantitative scale (e.g., from 0 to 10). The therapist should also be able to add the multimedia material related to each memory (photos, images, audios, or videos) and incorporate descriptive information of the stored material (e.g., location, description, date, and life cycle stage). The digital tool will allow the therapist to search for memories and consult a list of all recorded memories.

*Prepare a reminiscence therapy session.* To do this, the app will allow the therapist to view the patient's life story and attached reports, helping them to decide the nature of the session to be conducted. The practitioner can create the new reminiscence session by specifying a date, objectives, description and the life story memories the therapist wishes to work with, along with the related multimedia materials to be used.

*Conduct a reminiscence session.* During the session, the app will show the therapist all the data from the previously prepared session. During the session, the app should allow the practitioner to modify existing memories, incorporating new data obtained during the session, and to add memories not included previously in the patient's life story.

*End a reminiscence session.* At the end of a session, the therapist should be able to create a follow-up report, including the therapist's overall impression, the preservation of the memories worked with, and the emotional reaction that each memory evoked in the patient. The app should allow the therapist to consult all these reports at any time.

*Assess the patient.* The app should also allow the therapist to enter a periodic assessment of the patient, in order to have a record of the evolution of their condition. This clinical assessment will include relevant information such as diagnosis (e.g., type of dementia and date of diagnosis), stage of the condition (e.g., Global Deterioration Scale stage), results of standardized assessment scales, other observations of the therapist, and the overall impression of the patient's general condition (whether it has improved, worsened, or remains stable). Lastly, it should allow the therapist to consult these evaluations at any time and generate reports with their results.

*Automatically generate a life story.* The therapist should be able to generate multimedia narratives of life stories with videos/audios and images extracted from the patient's memories or the creation and printing of a book with the photos/images accompanying the life history events. This video or book should include a description of these memories. To create the life story, the therapist should be able to filter the events by type, date, and/or categories.

*Other requirements.* Other requirements suggested during this stage were: (1) the app should allow a patient to be assigned to several therapists; (2) it must be compatible with computers and tablets; (3) the app should be easy to use, regardless of the therapist's technical ability; (4) it should be efficient; it must save as much time as possible for the therapists; and (5) it should remain consistent, facilitating the learnability and use of the app.

*App prototype.* Once these requirements were clarified, the developers proceeded to design the first prototype. This design was then evaluated by the therapists of the research team. Table 2 presents the therapists' ratings of the degree of difficulty of the assessment tasks performed with the first prototype, on the SEQ tool. Most of the tasks (63.6%) received the maximum score (i.e., very easy). Only two tasks were scored below the mean response on the SEQ (5.3 and 5.6 points): ending a reminiscence session and writing an email to a patient.

Based on the results of the evaluation, the research team and developers agreed on the following changes: (1) clearer instructions for use and modified wording for clarity; the names of certain features of the app were changed: the follow-up reports were renamed "session reports", "types" of memories, was changed to "life stages"; (2) two new fields were added to the session creation form: barriers and facilitators; (3) the function to end a session was not intuitive, and a more prominent button was added; (4) in the clinical assessment form, a field was added for non-standardized assessment instruments used by the therapist; and (5) the option to sort and filter memories using different parameters was added (location, related persons, preservation status, and emotion the memory evokes in the patient). Once the design was agreed upon, the developers created a functional prototype that was evaluated in the second phase.

### Evaluation phase

#### Quantitative data

*SEQ instrument.* Table 3 shows the degree of difficulty of assessment tasks according to the therapists' responses to the SEQ. All tasks received scores between 6 and 7 points. Most tasks received a median score of 7 points (54.5%), i.e., the maximum possible score.

*SUS instrument.* On a scale from 0 to 100, the mean total score of the SUS questionnaire was 90.3 points (SD 9.68). Table 4 shows the scores for each of the SUS items.

*Potential benefits.* Regarding the potential benefits of using the app in reminiscence/life story therapy, the scores by therapists ranged from 4 to 5 points (medians), indicating that participants believed the digital tool would save practitioners time, enrich therapy sessions and improve their effectiveness (Table 5).

*Qualitative data.* Analyzing the responses to the open-ended questions of therapist interviews, two broad themes were established to summarize the qualitative results of the second phase: (a) acceptability of the app for reminiscence and life story therapy; and (b) areas for improvement.

*Acceptability of the app for reminiscence and life story therapy.* The therapists looked favorably on incorporating the *Recuerdame* app into therapy using reminiscence strategies. This theme encompasses three sub-themes: (1) views on positive features; (2) range of potential benefits; and (3) expectations of use.

*Views on positive features.* This sub-theme refers to the content and functionalities of the tool that therapists viewed positively. There was consensus among participants that the app was "*simple*", "*friendly*", "*intuitive*", and "*easy to use*". One therapist said that "*it is very comfortable to use*". Similarly, another participant explained: "*I had no doubts about using it*". Functionalities such as the automatic generation of personalized life stories using selected criteria or the possibility of easily locating and presenting stored multimedia content were positively valued by most of the participants, observing that these "*bring more dynamism to the therapies*" and "*enrich them [therapeutic interventions]*". Some therapists emphasized that the app enables the comprehensive management of a "*complex process*", consisting of "*a wide variety of tasks*": "*[the app] covers very well all the hard work that is done in reminiscence therapies*". The following excerpt from one participant illustrated this clearly: "*It's good that [the app] serves to address everything… from recording each patient's memories to creating life story books with one click*". Several therapists pointed out the possibilities that it offers for managing the task of retrieving information about patients' memories. For example, one participant noted that "*stored information can be retrieved at any time… easily and quickly*".

*Range of potential benefits.* All participants agreed that the app was potentially beneficial for therapists in reminiscence therapy. Comments on the perceived usefulness of the tool include: "*[the app] is very useful*"; "*it would help us a lot*". Most of the experts stressed that this app streamlines considerably the management of the therapeutic process in several ways: "*Although there is some initial work to enter all the data collected from the patient, but then the whole process to apply the therapy is much shorter*"; "*[The app] allows access to all the information needed during therapy in a very agile way*". Another commented that it also expedites clinical assessments of patients: "*for the [clinical] follow-up of each patient, it is not necessary to consult written documents, because all the information is digitized, and it is accessed very quickly*". Regarding the storage of information, some participants reported this functionality gave them feelings of "*security*" and "*trust*". This was supported by other qualitative comments: "*you keep everything in a safe place [i.e., in the app]*"; "*since the app is storing all the information that is incorporated, using it gives me confidence because I know that the saved information is not going to be lost*". One participant noted that digital storage of documents removes the need to keep materials related to memories: "*once [life story documents] are recorded in the app, they can be returned to the family… and I find it reassuring to see that nothing needs to be kept outside of it*". Many participants attached particular importance to the benefits of the automatic creation of life stories. For example, one therapist commented on the positive impact this can have on patients' self-confidence, viewing pleasant photos of their life story books and showing them to loved ones: "*it is very helpful to be able to print out the person's life story… we give them the life story book and the patients will proudly show it to their relatives, reinforcing their self-esteem and the value of their personal identity*". Another expert agreed with the potentially beneficial effect of sharing life story books with those closest to them: "*it is very positive that printed life story materials can be shared with people who are important to the patient*".

*Expectations of use.* This sub-theme shows the unanimous desire of the therapists to use the app in the future to facilitate their work using reminiscence and life story strategies. The following quotes illustrate this: "*Of course I would… I would use it on my patients daily*"; "*I would like to have this app at my workplace… to make better use of my work time*"; "*[the app] it has so many advantages that I would definitely use it in my work every day*".

*Areas for improvement.* This main theme captured the therapists' perceptions on areas for improvement of the prototype. Analysis revealed two sub-themes: (1) suggested changes to improve features and functionalities; and (2) recommendations on additional functionalities.

*Suggested changes to improve features and functionalities.* This sub-theme incorporated all the recommendations made by participants to improve the features and functionalities of the app. Some participants proposed improving the design and visual aspects, to better capture the attention of those with dementia and encourage their interest in the content and information:

"*I suggest making it [the app] more eye-catching at the perception level*".

"*The font size and the way the photos are presented are not very appealing to patients*".

"*The design could be improved, so that it can better capture patients' attention and encourage them to interact with all the possibilities offered by the app*".

Participants also commented that the app should provide more information on several key elements of therapy: (1) more patient information, adding, for example, file numbers, marital status, employment history, past and present leisure interests and hobbies; (2) more data on caregivers and related persons, including relevant information about them (for example, their profession), as well as "remarks" field to add additional details (for example, the characteristics of their relationship with the patient); and (3) more information about the therapy sessions, clarifying certain aspects such as the sequence of activities, location, a quantitative assessment of the degree of participation, therapist's judgment about the need to repeat the session, and proposals for improvements in future sessions. Finally, further suggestions included minor changes to the clinical assessment reports: adding a field for the therapist's signature; indicating the score ranges for each assessment instrument; and visualization of the patient's evolution using graphs, detailing the scores in assessment instruments in different clinical evaluations.

*Recommendations on additional functionalities.* Finally, this sub-theme included the participant's views on other functionalities to be added to the app. Two participants proposed adding features related to music, with the aim of incorporating its therapeutic potential into reminiscence strategies:

"*It would be very interesting to record in the app the favorite songs and singers of the person… Playing the songs of their lives would liven up the sessions and music always facilitates and favors the memories of their biographies*".

"*Undoubtedly, music is a great therapeutic ally in reminiscence therapies… It would be nice to link the app to websites like YouTube, where you can search for favorite songs from different stages of the patient's life and play them during therapeutic activities*".

Another suggested improvement was to add an internet search engine to "*search web pages for photographs of significant places in the lives of patients*", related to past moments and the present, to "*review the memories that these locations evoke*", as well as linking the app with the walk-through function available in Google Street View, which allows users to "*walk through streets, neighborhoods and places that were important in the person's life story… commenting on the patients' memories of where they lived and what they experienced in each place*".

## Discussion

The objective of this study was to develop the *Recuerdame* app, to help occupational therapists carry out reminiscence and life story therapy with older people with dementia. It is important to note that the process of reminiscence therapy is very time-consuming, involving not only the time allocated to sessions with patients, but also the time needed, both before and after the sessions, to collect and store abundant biographical information about each patient, planning objectives and preparing content, recording and monitoring the patient's evolution, etc..[29–31] Studies have shown that ICT tools have the potential to streamline a significant part of these tasks, particularly recording biographical materials and personalized memories, preparing structured intervention sessions or monitoring the progress of subjects.[13] The results of this study partially support our hypothesis, as occupational therapists generally agreed that *Recuerdame* could reduce therapy preparation time, enhance session delivery, and potentially improve overall effectiveness.

The participatory and user-centered approach adopted in designing the app encouraged the active involvement of future users throughout the entire process. As previous research has pointed out, this methodology is beneficial in several ways: designers generally lack specific knowledge of the needs in the field is compensated by the vision and broad experience of therapists; the quality and effectiveness of the technology is enhanced, with a closer connection between the app functionalities and the practical realities of the therapeutic tasks it is intended to support; there is less need for alterations and corrections in later phases of development; and, consequently, the tool finds greater acceptance by potential user.[32] In our study, an interdisciplinary research team, with a health sciences and computer science background, worked to co-develop a prototype. The therapists in the team determined the tasks that the app should facilitate and designed a functional tool that successfully incorporated all these tasks, which were quantitatively evaluated as easy to perform.

In the second phase of our research, we evaluated the perceptions of a sample of occupational therapists. The prototype was positively judged by participants: the vast majority of the functions of the tool were considered easy to perform by therapists (SEQ questionnaire), the level of usability of the tool was considered excellent according to the SUS questionnaire, and the quantitative data obtained indicated the acceptability of the app. One of the most popular features was the possibility of automatically obtaining multimedia narratives of life stories, with different formats and themes. The literature has highlighted that incorporating digital life story books and videos allows the development of an attractive multi-sensory environment especially conducive to cognitive stimulation, which can favor more meaningful therapeutic experiences for people with dementia, adapted to their individual profile of needs and interests.[8,33] Based on these findings, the *Recuerdame* app appears to be a versatile and innovative tool that empowers occupational therapists to manage reminiscence and life story interventions comprehensively, efficiently, and dynamically.

This work represents a novel exploration in Spain regarding the co-development of a digital application specifically designed to enhance reminiscence interventions for older people with dementia. Globally, research in this domain remains sparse, predominantly consisting of preliminary evaluations of the efficacy of diverse ICT systems in delivering personalized reminiscence therapies to those with dementia. Prior assessments have predominantly focused on aspects related to the functioning and well-being of older people with dementia, using case studies or pre-post designs with limited sample sizes.[8,12,14,15] In contrast, our research centers on supporting professionals in the field. We aimed not only to simplify their practice but also to diminish the demands on their time and resources. A distinctive contribution of this study lies in the development of a technological system designed to optimize the myriad tasks performed by occupational therapists. The co-designed application stands as a trailblazer, comprehensively addressing all aspects of implementing reminiscence therapy. The positive therapist feedback provides evidence of the potential benefits *Recuerdame* offers in areas such as collecting therapy materials, facilitating reminiscence sessions, and monitoring individual progress.

The identified functionalities in our application carry paramount importance due to their holistic approach, covering the entire spectrum of tasks involved in reminiscence therapy. This comprehensive nature distinguishes *Recuerdame* from existing tools, as it consolidates various functions into a singular, user-friendly platform. Comparatively, while other tools may serve specific functions individually, our co-creation process ensured that *Recuerdame* seamlessly integrates these functionalities. This integration is crucial in providing therapists with an all-encompassing tool that streamlines their workflow.

Furthermore, lessons learned from applying co-creation methodologies underscore the significance of collaboration with end-users. The iterative feedback loops and active involvement of occupational therapists in the development process not only enhanced the tool's functionalities but also ensured its practicality and user acceptance. The nuanced understanding gained through co-creation has implications beyond the development of *Recuerdame*, contributing to a broader understanding of effective co-creation in the context of therapeutic tools for dementia care.

In essence, our study introduces a novel technological solution (*Recuerdame*) and sheds light on the importance of comprehensive functionalities, distinctiveness from existing tools, and the insights gained from applying co-creation methodologies. We believe that these insights enrich the scientific literature in the field of reminiscence therapy for individuals with dementia.

Looking ahead, while well-received by occupational therapists, *Recuerdame* prompts further refinement, especially given its prototype status. Therapist feedback has been instrumental, identifying opportunities for feature improvement and offering recommendations. Future iterations aim to address these areas, embracing an iterative and collaborative approach to enhance overall effectiveness and user experience. Despite the thorough identification of functionalities during the co-design process, the dynamic nature of practical application revealed nuanced insights in the form of recommendations on additional functionalities during the evaluation phase. This prompts us to reconsider the synergy between co-design and evaluation. In subsequent iterations, our aim is to intensify the integration of insights garnered during the evaluation phase into the co-design process.

In this initial iteration of *Recuerdame*, our primary focus has centered on occupational therapists as the primary end-users, engaging intensively with the tool. However, it is crucial not to overlook the pivotal roles of individuals with dementia, who actively participate in therapy sessions, interacting with the tool by, for instance, viewing images crafted by therapists or watching videos capturing their life stories. Similarly, caregivers play a crucial role and stand to benefit from the tool, potentially using it at home to preserve and share memories with individuals with dementia. Our commitment extends to broadening the co-design process to include individuals with dementia and their caregivers, recognizing their integral roles. Ethical considerations for future iterations include implementing a robust and transparent consent process for individuals with dementia. Stringent measures will be in place to obtain explicit consent and ensure clear communication regarding data use, storage, and protection.

Despite these contributions, we acknowledge limitations. Firstly, saturation was reached after eight interviews which is a small sample size, although samples of this size are common in qualitative research.[34] Another limitation is that the study was limited to a particular region and the results may not be extrapolatable to all Spanish occupational therapy professionals in the field of care for people living with dementia. We recognize that the perception of this type of digital application may differ depending on the political, territorial or social context. On the other hand, participants were predominantly women. The composition of the sample was adjusted to the gender profile characteristic of the occupational therapy profession in Western countries.[35] However, previous studies have suggested there may be differences in the use and adaptation to digital tools between genders.[36] Finally, our research tested the application exclusively in an assessment setting, without using it in their professional practice. We consider it necessary to point out that there may be differences between our findings and the feedback provided by therapists when integrating the tool into their daily practice with the population of people with dementia. In addition, this study reflected the perceptions of occupational therapy experts. Consequently, future research should evaluate the use of this digital tool in professional practice settings and explore the perspectives of other healthcare professional profiles.

## Conclusions

Participating occupational therapists generally agreed that the technological solution co-designed in this study appears to be a versatile tool that empowers these professionals to manage reminiscence interventions. The study findings suggested the usability and acceptability of *Recuerdame*. As end-users, occupational therapists were overwhelmingly positive about the use of this digital tool in reminiscence interventions with older people with dementia. Future research should refine the app, addressing opportunities for improvement identified by participants, through an iterative and collaborative approach. In addition, more research is needed in this scientific field with the objectives of including people with dementia and their caregivers in the co-design process, learning about the use of the app in professional practice settings, and analyzing the perspectives of other healthcare professionals.

**Acknowledgements**: The authors thank all the therapists who participated in the evaluation phase for their time and contributions to this research. Additionally, we thank the students Cristina Barquilla, Patricia Díez, Santiago Marco Mulas and Eva Verdú for their contribution to the prototypes.

**Contributorship:** All authors conceived the study. IDC and AAA were involved in protocol development, gaining ethical approval, participant recruitment and data analysis. VFG and SBB were involved in protocol development and data collection. IDC wrote the manuscript as the first author. All authors read, reviewed, and approved the final version of the manuscript.

**Declaration of conflicting interests:** The authors declared no potential conflicts of interest with respect to the research, authorship, and/or publication of this article.

**Ethical approval:** The Research Ethics Committee of the University of A Coruña (Spain) approved this study (identification code 2022-024).

**Funding:** The authors disclosed receipt of the following financial support for the research, authorship, and/or publication of this article: This work was supported by the Spanish Ministry of Science and Innovation, (grant number PID2019-108927RB-I00).

**Guarantor:** IDC

**ORCID iD:** Iván De-Rosende-Celeiro  https://orcid.org/0000-0002-4569-2050

**Supplemental material:** Supplemental material for this article is available online.

## References

1. Alzheimer's Association. 2022 Alzheimer's disease facts and figures. *Alzheimers Dement J Alzheimers Assoc* 2022; 18: 700–789.
2. World Health Organization. *Global status report on the public health response to dementia*. Geneva: World Health Organization, 2021.
3. Cammisuli DM, Danti S, Bosinelli F, et al. Non-pharmacological interventions for people with Alzheimer's disease: a critical review of the scientific literature from the last ten years. *Eur Geriatr Med* 2016; 7: 57–64.
4. Woods B, O'Philbin L, Farrell EM, et al. Reminiscence therapy for dementia. *Cochrane Database Syst Rev* 2018; 3: CD001120.
5. Dempsey L, Murphy K, Cooney A, et al. Reminiscence in dementia: a concept analysis. *Dement Lond Engl* 2014; 13: 176–192.

6. Morris RG. Recent developments in the neuropsychology of dementia. *Int Rev Psychiatry* 1994; 6: 85–107.
7. Schweitzer P. Reminiscence in dementia care. *Int J Reminisc Life Rev* 2013; 1: 42–47.
8. Cuevas PEG, Davidson PM, Mejilla JL, et al. Reminiscence therapy for older adults with Alzheimer's disease: a literature review. *Int J Ment Health Nurs* 2020; 29: 364–371.
9. Saragih ID, Tonapa SI, Yao CT, et al. Effects of reminiscence therapy in people with dementia: a systematic review and meta-analysis. *J Psychiatr Ment Health Nurs* 2022; 29: 883–903.
10. Park K, Lee S, Yang J, et al. A systematic review and meta-analysis on the effect of reminiscence therapy for people with dementia. *Int Psychogeriatr* 2019; 31: 1581–1597.
11. Elfrink TR, Zuidema SU, Kunz M, et al. Life story books for people with dementia: a systematic review. *Int Psychogeriatr* 2018; 30: 1797–1811.
12. Subramaniam P and Woods B. Towards the therapeutic use of information and communication technology in reminiscence work for people with dementia: a systematic review. *Int J Comput Healthc* 2010; 1: 106–125.
13. Lazar A, Thompson H and Demiris G. A systematic review of the use of technology for reminiscence therapy. *Health Educ Behav Off Publ Soc Public Health Educ* 2014; 41: 51S–61S.
14. Cho E, Shin J, Seok JW, et al. The effectiveness of non-pharmacological interventions using information and communication technologies for behavioral and psychological symptoms of dementia: a systematic review and meta-analysis. *Int J Nurs Stud* 2022; 138: 104392.
15. Dequanter S, Gagnon MP, Ndiaye MA, et al. The effectiveness of e-health solutions for aging with cognitive impairment: a systematic review. *Gerontologist* 2021; 61: e373–e394.
16. Bano M and Zowghi D. User involvement in software development and system success: a systematic literature review. In: *Proceedings of the 17th International Conference on Evaluation and Assessment in Software Engineering [Internet]*, New York, NY, USA: Association for Computing Machinery. 2013 [cited 2023 Jan 11]. p. 125–130. (EASE '13). Available from: https://doi.org/10.1145/2460999.2461017
17. Kujala S. User involvement: a review of the benefits and challenges. *Behav Inf Technol* 2003; 22: 1–16.
18. Bridges J, Gray W, Box G, et al. Discovery interviews: a mechanism for user involvement. *Int J Older People Nurs* 2008; 3: 206–210.
19. Lewis JR. Usability testing. In: Salvendy G (eds) *Handbook of human factors and ergonomics*. 4th ed. New York, NY: John Wiley & Sons, 2012, pp.1267–1312.
20. Lewis JR. Usability: lessons learned … and yet to be learned. *Int J Human–Computer Interact* 2014; 30: 663–684.
21. Sauro J and Lewis JR. *Quantifying the User Experience: Practical Statistics for User Research*. 2nd ed. Cambridge, MA: Elsevier / Morgan-Kaufmann, 2016.
22. Carpenter C and Suto M. *Qualitative research for occupational and physical therapists: a practical guide [Internet]*. Oxford, UK: Wiley, 2008, Available from: http://eu.wiley.com/WileyCDA/WileyTitle/productCd-1405144351.html.
23. Sevilla-Gonzalez MDR, Moreno Loaeza L, Lazaro-Carrera LS, et al. Spanish version of the system usability scale for the assessment of electronic tools: development and validation. *JMIR Hum Factors* 2020; 7: e21161.
24. World Medical Association. World Medical Association Declaration of Helsinki: ethical principles for medical research involving human subjects. *JAMA* 2013; 310: 2191–2194.
25. Braun V and Clarke V. Using thematic analysis in psychology. *Qual Res Psychol* 2006; 3: 77–101.
26. Green J and Thorogood N. *Qualitative Methods for Health Research*. 3rd ed. London: Sage, 2014.
27. Graneheim UH and Lundman B. Qualitative content analysis in nursing research: concepts, procedures and measures to achieve trustworthiness. *Nurse Educ Today* 2004; 24: 105–112.
28. Tong A, Sainsbury P and Craig J. Consolidated criteria for reporting qualitative research (COREQ): a 32-item checklist for interviews and focus groups. *Int J Qual Health Care* 2007; 19: 349–357.
29. Gowans G, Campbell J, Alm N, et al. Designing a multimedia conversation aid for reminiscence therapy in dementia care environments. In: *CHI '04 extended abstracts on human factors in computing systems [internet]*. New York, NY, USA: Association for Computing Machinery, 2004 [cited 2023 Jan 9], pp.825–836. (CHI EA '04). Available from: https://doi.org/10.1145/985921.985943
30. Paay J, Kjeldskov J, Aaen I, et al. User-centred iterative design of a smartwatch system supporting spontaneous reminiscence therapy for people living with dementia. *Health Informatics J* 2022; 28: 14604582221106002.
31. Moon S and Park K. The effect of digital reminiscence therapy on people with dementia: a pilot randomized controlled trial. *BMC Geriatr* 2020; 20: 166.
32. Damodaran L. User involvement in the systems design process-a practical guide for users. *Behav Inf Technol* 1996; 15: 363–377.
33. Yasuda K, Kuwabara K, Kuwahara N, et al. Effectiveness of personalised reminiscence photo videos for individuals with dementia. *Neuropsychol Rehabil* 2009; 19: 603–619.
34. Guest G, Bunce A and Johnson L. How many interviews are enough?: an experiment with data saturation and variability. *Field Methods* 2006; 18: 59–82.
35. Ledgerd R. World federation of occupational therapists. WFOT report: WFOT human resources project 2018 and 2020. *World Fed Occup Ther Bull* 2020; 76: 69–74.
36. Goswami A and Dutta S. Gender differences in technology usage—A literature review. *Open J Bus Manag* 2016; 4: 51–59.